# THE BUSINESS MODEL BANK: CONCEPTUALIZING A DATABASE STRUCTURE FOR LARGE-SAMPLE STUDY OF AN EMERGING MANAGEMENT CONCEPT

Fredrik Hacklin[1], Nobuaki Minato[2] and Toma Kobayashi[2]

[1]ETH Zurich
Department of Management, Technology and Economics, Zurich, 8092, Switzerland
e-mail: fhacklin@ethz.ch

[2]Keio University
Graduate School of System Design and Management
Yokohama, 223-8526, Japan
e-mail: minato.nobuaki@sdm.keio.ac.jp

## ABSTRACT

The 'business model' represents an increasingly important management concept. However, progress in research related to the concept is currently inhibited from inconsistencies in terms of formalizing and therewith also empirically measuring the 'business model' concept. Taking this as a starting point, this paper offers a conceptualization for building a scalable database to rigorously capture large samples of business models. The following contributions are made: First, we suggest a concept for dimensions to be modeled in the database. Second, we discuss issues critical to the scalability of such an endeavor. Third, we point to empirical and simulation-based studies enabled by the population of such a database. Considerations for theory and practice are offered.

## INTRODUCTION

The concept of the 'business model' (BM) represents an increasingly contested and widely adopted analytic lens for management scholars and practitioners (Zott, Amit, & Massa, 2011; Osterwalder & Pigneur, 2010). While there exists broad consensus on the concept's strategic importance for modeling how a focal firm creates and captures value within its ecosystem of exchange partners (e.g. Zott & Amit, 2010), the concept has been surprisingly little operationalized as a unit of analysis in empirical studies, and if so, oftentimes in mutually inconsistent manner. This may be due to a lacking consensus on a commonly formalized definition in the scholarly literature (Klang, Wallnöfer, & Hacklin, 2014). As a result, much of the empirical research around the BM rests on anecdotal evidence, or primary data collection that turns out hard to replicate.

Against this background, this paper explores opportunities for creating a global database for collecting BM data along a rigorous set of dimensions, allowing the data collection to be scaled across multiple research teams, countries and empirical contexts. Through populating a comprehensive database either through primary or secondary data, this offers the ground for cataloguing BMs across, for example, different industries, regions, and firms sizes. We contend that such a 'business model bank' would bear the potential to not only provide a much-needed basis for conducting larger-scale empirical studies, but moreover could it also cross-fertilize the debate around the lacking consensus on an actionable format of the BM.

This paper is organized as follows: First, we introduce and describe the concept of the BM database. Second, we highlight critical considerations to the feasibility of such a database furthering rigorous empirical research. Finally, we discuss our findings and point to avenues for further research.

## THE 'BUSINESS MODEL BANK' CONCEPT

Given the large variety of definitions and frameworks suggested to formalize the BM, recent attempts to synthesize these into one commonly unified meta-framework have largely failed, or resulted in yet another novel definition (for a discussion, cf. Klang et al., 2014). At the same time, the BM concept can be regarded as one of the most popular management concepts, both within the scholarly discourse, as well as among management practitioners (cf. Baden-Fuller & Morgan, 2010). In response to these resulting 'paradoxical tensions' (Klang et al., 2014), we intend to create the 'business model bank' to further empirical research, thereby contributing to strengthening the concept. The database is described in the following.

### Concept and Dimensions

Figure 1 illustrates the overview of the *Business Model Bank (BMB)* concept. We develop a database in which multiple, BMs can be collected and stored. The structure and hierarchy follows a simplified formalization of the BM, which rests on the BM canvas as developed by Osterwalder & Pigneur (2010), yet complemented by structural elements by the top 10[1] most cited academic papers suggesting frameworks for BMs. Variables modeled in the database are either operationalizations following the BM canvas, or they represent established proxies extracted from previous academic studies. The implementation will follow the structural template of a relational database, allowing for a hierarchical modeling of database elements. The primary key of a given record will be the underlying firm. Table 1 shows the tentative dimensions modeled in the database. These dimensions are based on a cursory review of conceptual contributions in the BM literature.

---

[1] These are omitted here for reasons of space. For an extensive review, see Zott, Amit & Massa (2011).

*Table 1: Dimensions of BMB Database*

| Dimension | Key Information |
| --- | --- |
| Industry | Domain (ISIC, rev.4) |
| | Size |
| | Growth Rate |
| Player | Function |
| | Number |
| Exchange (among players) | Product |
| | Service |
| | Information |
| | Cash |
| Infrastructure (for exchange) | Transportation |
| | Location |
| | Communication |
| | Transaction |
| Resource | Financial resources |
| | Human resources |
| | Natural resources |
| | Intellectual Resources |

In addition to the structural design of the database, the inclusion of an additional market-based feature for accelerating the adoption of BMB will be considered.[2] Encountering both classic and novel BM ideas would in such a scenario represents the key value proposition of the BMB.

**Scalability**

In order for the implementation of the BMB database to scale, the following critical considerations need to be made. First, the dimensions need to be well rooted in peer-reviewed academic literature. This is particularly important in order for the BMB database to result in further empirical studies with significant impact. Second, dimensions need to be mutually exclusive, and ideally, collectively exhaustive. Without the mutual exclusivity, data collection may in practice turn out ambiguous, as it may not be clear which category to place a given observation into. Finally, we need to have a clearly formulated and unambiguous data collection protocol, allowing the data collection to be consistent, even though scaled between different contextual settings (e.g., among various research groups and involved students), in order to guarantee full transparency and comparability of results. In this regard, our proposition is to provide a handbook for the BMB, which specifies the rules to follow for collecting and coding the data.

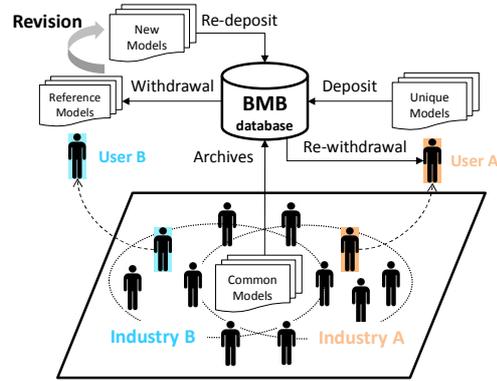

*Figure 1: BMB Concept*

**DISCUSSION**

Through the construction of the BMB database, we are responding to the lack of consensus on the BM construct on the one hand, as well as to the currently rather poor level of replication in empirical research around the BM on the other. To remedy these shortcomings, the BMB provides the basis for large-scale empirical studies, allowing, for example, to investigate the link between changes in firms' BMs and their performance on the market, for example through conducting quantitative analysis on a large sample of firms and their business models. This type of further research efforts would contribute to closing the gap in terms of the 'paradoxical tensions' as highlighted by Klang et al., (2014).

---

[2] Consider the following scenario: Unanimous users can deposit unique BMs as valuable assets instead of cash. The users are also allowed withdraw BMs in the stock. Thus, interest is paid by not cash but a variety of unfamiliar BMs, which belong to other industries.